# Instruction Set Architecture (ISA) for Processing-in-Memory DNN Accelerators

Xiaoming Chen
*Institute of Computing Technology, Chinese Academy of Sciences*
Email: chenxiaoming@ict.ac.cn

**Abstract** ---- In this article, we introduce an instruction set architecture (ISA) for processing-in-memory (PIM) based deep neural network (DNN) accelerators. The proposed ISA is for DNN inference on PIM-based architectures. It is assumed that the weights have been trained and programmed into PIM-based DNN accelerators before inference, and they are fixed during inference. We do not restrict the devices of PIM-based DNN accelerators. Popular devices used to build PIM-based DNN accelerators include resistive random-access memory (RRAM), flash, ferroelectric field-effect transistor (FeFET), static random-access memory (SRAM), etc. The target DNNs include convolutional neural networks (CNNs) and multi-layer perceptrons (MLPs). The proposed ISA is transparent to both applications and hardware implementations. It enables to develop unified toolchains for PIM-based DNN accelerators and software stacks. For practical hardware that uses a different ISA, the generated instructions by unified toolchains can easily converted to the target ISA. The proposed ISA has been used in the open-source DNN compiler PIMCOMP-NN (https://github.com/sunxt99/PIMCOMP-NN) and the associated open-source simulator PIMSIM-NN (https://github.com/wangxy-2000/pimsim-nn).

## 1. Architecture Abstraction

Figure 1 shows the architecture overview. In general, a DNN acceleration system is comprised of a set of cores and a global memory. The cores are connected by some sort of interconnection. We do not restrict the detailed implementations of the cores, memory, and interconnection. The cores can be in a single chip, or separate chiplets, etc., and the interconnection can be a shared bus, a network-on-chip (NoC), or a mesh, etc. The global memory is shared by all cores. The weights of DNNs are stored in the cores, while the inputs are stored in the global memory.

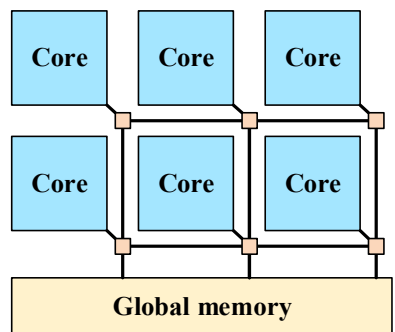

Figure 1: Abstract architecture.

Figure 2 shows the abstracted core architecture. Each core has its own instruction sequence and instruction decoder. Different cores execute different instructions asynchronously. A core has three main execution units: a scalar unit, a PIM matrix unit and a vector unit, and three main memories: an instruction memory, a register file and a local memory. The scalar unit and the register file are used for executing register/scalar instructions, which are for addressing the local memory and the global memory. The local memory works as a scratchpad memory that stores DNN's data, including the primary inputs/outputs and the inputs/outputs of the layers. The PIM matrix unit executes matrix-vector multiplications and the vector unit executes other DNN operations including ReLU, pooling, sigmoid, etc. In practical implementations, general-purpose processors (e.g., RISC-V cores) may provide the functions of the scalar and vector units, and the PIM matrix unit is typically realized by crossbar arrays. The PIM matrix unit and the vector unit can only access the local memory, with addresses from the register file and/or the scalar unit. The kernel components of the PIM matrix unit and the vector unit may work in the analog domain but the interfaces are purely in the digital domain.

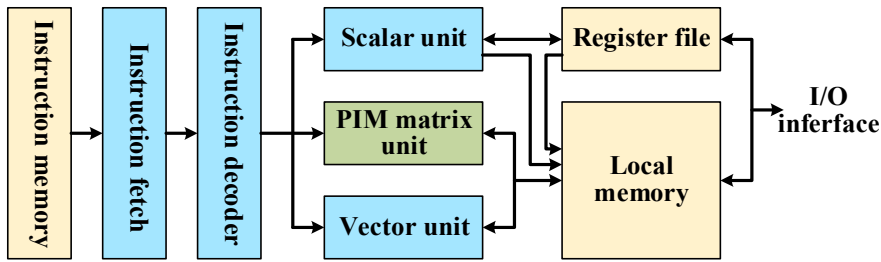

Figure 2: Abstract core architecture.

## 2. DNN Mapping

We provide a general description of the mapping of one or more DNNs' inference onto the above architecture. Basically, a core can map multiple layers, and a layer can be mapped to multiple cores. The software toolchain should optimize the mapping of the layers onto the cores, and generate the mapping information, which is read by the hardware for programming. The hardware will store the mapping information which will also be used for matrix-vector multiplications in inference. In a core, matrix-vector multiplications are executed by the PIM matrix unit, and other operations like ReLU, pooling, sigmoid, etc., are executed by the vector unit.

A core has multiple arrays (e.g., RRAM-based crossbar arrays). Here an array means a logical array that can complete a matrix-vector multiplication in one step. A logical array can be composed of multiple physical arrays. For example, two arrays are needed to store the positive and negative weights, respectively. In this case, the two arrays are treated as a logical array. For another example, if the device precision is 4 bits and the weight precision is 8 bits, then the needed physical arrays are doubled. The ISA and the toolchain are only concerned with logical arrays. Physical arrays are transparent to software.

The arrays in a core are numbered from 0. In the proposed ISA, the concept of *array group* is involved. An array group can include an arbitrary number of arrays at arbitrary positions in a core. For example, the arrays that are for the same layer can be organized as an array group. Array groups in a core are numbered from 0. The organization of array groups is generated by the toolchain. The hardware will store the array group information. The introduction of array groups if for the convenience of matrix-vector multiplications. Matrix-vector multiplications are implemented at the granularity of an array group. For a matrix-vector multiplication, the array group index with the input vector address and the output vector address is provided, so that the input vector can be multiplied with the matrix specified by the array group index.

Ref. [1] provides a technical description of the DNN mapping methodology for PIM architectures.

## 3. ISA Design Principles

- The instructions are high-level abstractions of DNN's primary operators, focusing on matrix and vector operations. The features of both DNN's operators and hardware primitives are taken into account when designing the ISA. The instructions will reflect practical operations of DNNs, while the low-level hardware implementations are transparent to software. This is for the universality of the ISA.
- The ISA supports different bit-widths for different layers. The input bit-width and the output bit-width are set by a common instruction.
- The instructions are designed to be as neat as possible, avoiding special formats of some instructions.
- By taking into account the features of vector operations, an elaborate offset format is introduced.
- Due to the determinacy of operations and addresses, the ISA uses immediate values as many as possible, minimizing register operations. In addition, there is no branch instruction, so loops should be completely unrolled.
- The control path and the data path are completely separated. The registers are only used for addressing and control, and the local memories are only used for storing DNNs' data.
- The ISA is for standalone DNN accelerators. It has no relation with any CPU's ISA.
- The ISA only contains essential instructions for DNNs' inference on PIM architectures. It is not complete for DNNs' training or even general-purpose computing.

## 4. Instruction Definition

### 4.1 Convention
(1) The proposed ISA is 32-bit or 64-bit fixed-length instructions.
(2) Registers are of size 32 bits. The ISA supports 32 registers at most.
(3) Global memory address is 64 bits which needs two consecutive registers starting from an even register index.
(4) Registers are only used for addressing, and local memories are only used for storing DNN's data.
(5) Registers are denoted as the form of $rid. Specifically, $rd denotes the destination register, and $rs1 and $rs2 denote the source registers.
(6) Local memory address and global memory address are denoted as LMem[addr] and GMem[addr], respectively, or LMem[start, end] and GMem[start, end], where end can be expressed as ~+size-1 (~ denotes the corresponding start address which is omitted). Here 'addr', 'start', 'end' and 'size' are all expressed in bytes.
(7) imm or imm_xxx denotes an immediate value.
(8) Many instructions have an offset field. Offset has two sub-fields: offset_select and offset_value. Offset_select has 3 bits that indicate whether $rd, $rs1 and $rs2 will be added with the address offset

to get the final address. Among them, offset_select[0] is for $rd, offset_select[1] is for $rs1 and offset_select[2] is for $rs2. Offset_value that denotes the element offset or byte offset is a single value for all selected registers. For vector instructions, offset_value is an element offset, and it will be multiplied with the elementary byte-width to get the byte offset. For other instructions, it is a byte offset. If an instruction only has the offset_value field, then the corresponding byte offset is directly added to the specified register value without the register selection mechanism.

(9) Matrix/vector elements must be aligned at bytes in local memories. For example, if a matrix element is of size 10 bits, then each element uses the 10 low-order bits of 2 bytes.

(10) Different layers can use different input bit-widths and different output bit-widths.

(11) Each core keeps the bit-widths and byte-widths of vector inputs and outputs for calculating the addresses of matrix/vector elements, which are denoted by ibiw (bit-width of inputs), obiw (bit-width of outputs), ibyw (byte-width of inputs), obyw (byte-width of outputs), respectively.

## 4.2 Scalar/register Instructions

(1) **sldi** $rd, imm
Set register immediate value: $rd = imm, where imm is of size 4 bytes.

(2) **sld** $rd, $rs1, offset_byte
Load a 4-byte value from global memory to register: $rd = GMem[$rs1+offset_byte, ~+3]. rs1 must be an even register index.

(3) **sadd** $rd, $rs1, $rs2
Integer addition: $rd = $rs1 + $rs2.

(4) **ssub** $rd, $rs1, $rs2
Integer subtraction: $rd = $rs1 - $rs2.

(5) **smul** $rd, $rs1, $rs2
Integer multiplication: $rd = $rs1 * $rs2.

(6) **saddi** $rd, $rs1, imm
Integer addition with immediate value: $rd = $rs1 + imm. There is no subi instruction as it can be implemented by saddi.

(7) **smuli** $rd, $rs1, imm
Integer multiplication with immediate value: $rd = $rs1 * imm.

## 4.3 Matrix/vector Instructions

(1) **setbw** ibiw, obiw
Set the bit-widths of each element for input vectors and output vectors. Related vector instructions use the configured bit-widths. Once setbw is caled, all subsequent related vector instructions will use the configured bit-widths, until a new setbw is called. Once ibiw and obiw are set, ibyw and obyw are also set accordingly by the hardware.

If the hardware does not support variable bit-width, this instruction is invalid and the matrix/vector instructions use the fixed bit-width of the hardware.

(2) **mvmul** $rd, $rs1, mbiw, imm_relu, imm_group

Matrix-vector multiplication: imm_group specifies the array group which is generated by the toolchain before inference. The input vector starts at address LMem[$rs1] and its length must match the corresponding matrix dimension of the array group. The output vector is written to address LMem[$rd] and its length is determined by the corresponding matrix dimension of the array group. imm_relu is 1 bit that specifies whether a ReLU operation will be applied for the resulting vector. The input vector bit-width is subject to ibiw, the input matrix bit-width is subject to mbiw, and the output vector bit-width is subject to obiw.

(3) **vvadd** $rd, $rs1, $rs2, imm_len, offset

Vector-vector addition: vector starting at address LMem[$rs1+ibyw*offset_value*offset_select[1]] plus vector starting at address LMem[$rs2+ibyw*offset_value*offset_select[2]], both of length imm_len, the summation vector is written to address LMem[$rd+ibyw*offset_value*offset_select[0]]. The input bit-width and output bit-width are both subject to ibiw.

(4) **vvsub** $rd, $rs1, $rs2, imm_len, offset

Vector-vector subtraction: vector starting at address LMem[$rs1+ibyw*offset_value*offset_select[1]] minus vector starting at address LMem[$rs2+ibyw*offset_value*offset_select[2]], both of length imm_len, and the difference vector is written to address LMem[$rd+ibyw*offset_value*offset_select[0]]. The input bit-width and output bit-width are both subject to ibiw.

(5) **vvmul** $rd, $rs1, $rs2, imm_len, offset

Element-wise vector-vector multiplication: vector starting at address LMem[$rs1+ibyw*offset_value*offset_select[1]] times vector starting at address LMem[$rs2+ibyw*offset_value*offset_select[2]], both of length imm_len, and the product vector is written to address LMem[$rd+obyw*offset_value*offset_select[0]]. The input bit-width is subject to ibiw and the output bit-width is subject to obiw.

(6) **vvdmul** $rd, $rs1, $rs2, imm_len, offset

Vector-vector dot multiplication (inner product): vector starting at address LMem[$rs1+ibyw*offset_value*offset_select[1]] dot-times vector starting at address LMem[$rs2+ibyw*offset_value*offset_select[2]], both of length imm_len, the dot product result (one element) is written to address LMem[$rd] (offset_select[0] does not apply). The input bit-width is subject to ibiw and the output bit-width is subject to obiw.

(7) **vvmax** $rd, $rs1, $rs2, imm_len, offset

Element-wise vector-vector maximum: vector starting at address LMem[$rs1+ibyw*offset_value*offset_select[1]] compared with vector starting at address LMem[$rs2+ibyw*offset_value*offset_seelct[2]], both of length imm_len, the element-wise maximum vector is written to address LMem[$rd+ibyw*offset_value*offset_select[0]]. The input

bit-width and output bit-width are both subject to ibiw.

(8) **vvsll** $rd, $rs1, $rs2, imm_len, offset

Element-wise shift-left logic: elements of vector starting at address LMem[$rs1+ibyw*offset_value*offset_select[1]] are logically left shifted according to the number of bits specified in the elements of vector starting at address LMem[$rs2+ibyw*offset_value*offset_select[2]], both of length imm_len, and the resulting vector is written to address LMem[$rd+obyw*offset_value*offset_select[0]]. The input bit-width is subject to ibiw and the output bit-width is subject to obiw.

(9) **vvsra** $rd, $rs1, $rs2, imm_len, offset

Element-wise shift-right arithmetic: elements of vector starting at address LMem[$rs1+ibyw*offset_value*offset_select[1]] are arithmetically right shifted according to the number of bits specified in the elements of vector starting at address LMem[$rs2+ibyw*offset_value*offset_select[2]], both of length imm_len, and the resulting vector is written to address LMem[$rd+obyw*offset_value*offset_select[0]]. The input bit-width is subject to ibiw and the output bit-width is subject to obiw.

(10) **vavg** $rd, $rs1, $rs2, imm_len, offset_value

Vector average: calculating the average of vector of length imm_len starting at address LMem[$rs1+ibyw*offset_value], with stride size $rs2 (elements), and the result (one element) is written to address LMem[$rd]. The input bit-width is subject to ibiw and the output bit-width is subject to obiw.

(11) **vrelu** $rd, $rs1, imm_len, offset

Vector relu: calculating relu on vector of length imm_len starting at address LMem[$rs1+ibyw*offset_value*offset_select[1]], and the resulting vector is written to address LMem[$rd+ibyw*offset_value*offset_select[0]]. The input bit-width and output bit-width are both subject to ibiw.

(12) **vtanh** $rd, $rs1, imm_len, offset

Vector tanh: calculating tanh on vector of length imm_len starting at address LMem[$rs1+ibyw*offset_value*offset_select[1]], and the resulting vector is written to address LMem[$rd+obyw*offset_value*offset_select[0]]. The input bit-width is subject to ibiw and the output bit-width is subject to obiw.

(13) **vsigm** $rd, $rs1, imm_len, offset

Vector sigmoid: calculating sigmoid on vector of length imm_len starting at address LMem[$rs1+ibyw*offset_value*offset_select[1]], and the resulting vector is written to address LMem[$rd+obyw*offset_value*offset_select[0]]. The input bit-width is subject to ibiw and the output bit-width is subject to obiw.

(14) **vmv** $rd, $rs1, $rs2, imm_len

Vector move: gathering strided elements to a consecutive vector. The start address of destination vector is LMem[$rd], and the start address of source data is LMem[$rs1]. The stride size is $rs2 (elements). The number of elements to be gathered is imm_len. If the source data and the destination data have overlap, the result of the overlapped area is undefined. The input bit-width and output bit-width are both subject to ibiw.

(15) **vrsu** $rd, $rs1, $rs2, imm_len, offset
Element-wise vector resize with upper bound: resizing elements of vector starting at address LMem[$rs1+ibyw*offset_value*offset_select[1]] of length imm_len to vector starting at address LMem[$rd+obyw*offset_value*offset_select[0]]. The output element values will be clamped at the upper bound specified in $rs2. This means that if an original value is larger than $rs2, the value is set to $rs2. The input bit-width is subject to ibiw and the output bit-width is subject to obiw. If the byte-width does not increase, input data and output data may have overlap; otherwise cannot.
If the hardware does not support variable bit-width, this instruction is invalid.

(16) **vrsl** $rd, $rs1, $rs2, imm_len, offset
Element-wise vector resize with lower bound: resizing elements of vector starting at address LMem[$rs1+ibyw*offset_value*offset_select[1]] of length imm_len to vector starting at address LMem[$rd+obyw*offset_value*offset_select[0]]. The output element values will be clamped at the lower bound specified in $rs2. This means that if an original value is smaller than $rs2, the value is set to $rs2. The input bit-width is subject to ibiw and the output bit-width is subject to obiw. If the byte-width does not increase, input data and output data may have overlap; otherwise cannot.
If the hardware does not support variable bit-width, this instruction is invalid.

### 4.4 Communication/synchronization Instructions

(1) **ld** $rd, $rs1, imm_size, offset
Load data from global memory to local memory: LMem[$rd+offset_byte*offset_select[0], ~+imm_size-1] = GMem[$rs1+offset_byte*offset_select[1], ~+imm_size-1]. rs1 must be an even register index.

(2) **st** $rd, $rs1, imm_size, offset
Store data from local memory to global memory: GMem[$rd+offset_byte*offset_select[0], ~+imm_size -1] = LMem[$rs1+offset_byte*offset_select[1], ~+imm_size-1]. rd must be an even register index.

(3) **lldi** $rd, imm, imm_size, offset_byte
Set local memory immediate value: LMem[$rd+offset_byte, ~+imm_size-1] = imm (repeatedly setting imm_size bytes to imm). The size of imm is 1 byte. This instruction behaves similar to the C runtime function "memset".

(4) **lmv** $rd, $rs1, imm_size, offset
Move local memory data: LMem[$rd+offset_byte*offset_select[0], ~+imm_size-1] = LMem[$rs1+offset_byte*offset_select[1], ~+imm_size-1]. After data moving, the source data is not changed or cleared. If the source data and the destination data have overlap, the result of the overlapped area is undefined. This instruction behaves similar to the C runtime function "memcpy".

(5) **send** $rs1, imm_core, imm_size, offset_byte

Send source data from LMem[$rs1+offset_byte, ~+imm_size-1] to the local memory of destination core imm_core. This instruction is synchronous (only when the data is received by the destination core, it returns). The toolchain must ensure that the data sending and receiving operations are successful (there is a matched recv instruction invoked on a different core, and the receiving buffer is sufficient). If the architecture does not support core-to-core data transmission, this instruction is disabled. In that case, the global memory can be used for core-to-core data transmission.

(6) **recv** $rd, imm_core, imm_size, offset_byte

Receive data from core imm_core, storing received data to LMem[$rd+offset_byte, ~+imm_size-1]. This instruction is synchronous (only when the data is fully received, it returns). It must match a send instruction invoked on a different core. If the architecture does not support core-to-core data transmission, this instruction is disabled. In that case, the global memory can be used for core-to-core data transmission.

(7) **wait** imm_ev imm_val

This instruction is used together with sync to realize a simple synchronization mechanism. Each core has a few event registers to store event values which are reset to zero when powered up. The wait instruction checks whether the value of event register imm_ev is imm_val. If not, the calling core enters a blocked state until the value of event register imm_ev becomes imm_val. Once the target value becomes imm_val, it is reset to zero and the calling core continues executing subsequent instructions. Check and reset are executed in an atomic way.

The synchronization mechanism provided by wait and sync does not guarantee the absolute correctness or deadlock-free in any synchronization situations. The toolchain should generate correct code for synchronization.

(8) **sync** imm_ev imm_core

This instruction increases the event register indexed imm_ev of core imm_core by 1. The operation is atomic and synchronous.

## 5. Instructions Summary

Figure 3 summarizes the instructions. The length of the defined instructions can be either 32-bit or 64-bit. In the 32-bit mode, the offset mechanism is not supported, and for sldi, saddi and smuli, the 'imm' field is in the unused field of the lower 32 bits.

Notes:
- The register id colored blue must be an even number.
- The unused area in the offset field is 'offset_select', which occupies 3 bits.
- The instructions and fields colored red are used for hardware that supports variable bit-width.

## References

[1] Xiaotian Sun, Xinyu Wang, Wanqian Li, Lei Wang, Yinhe Han, Xiaoming Chen, "PIMCOMP: A Universal Compilation Framework for Crossbar-based PIM DNN Accelerators", in Design Automation Conference (DAC'23).

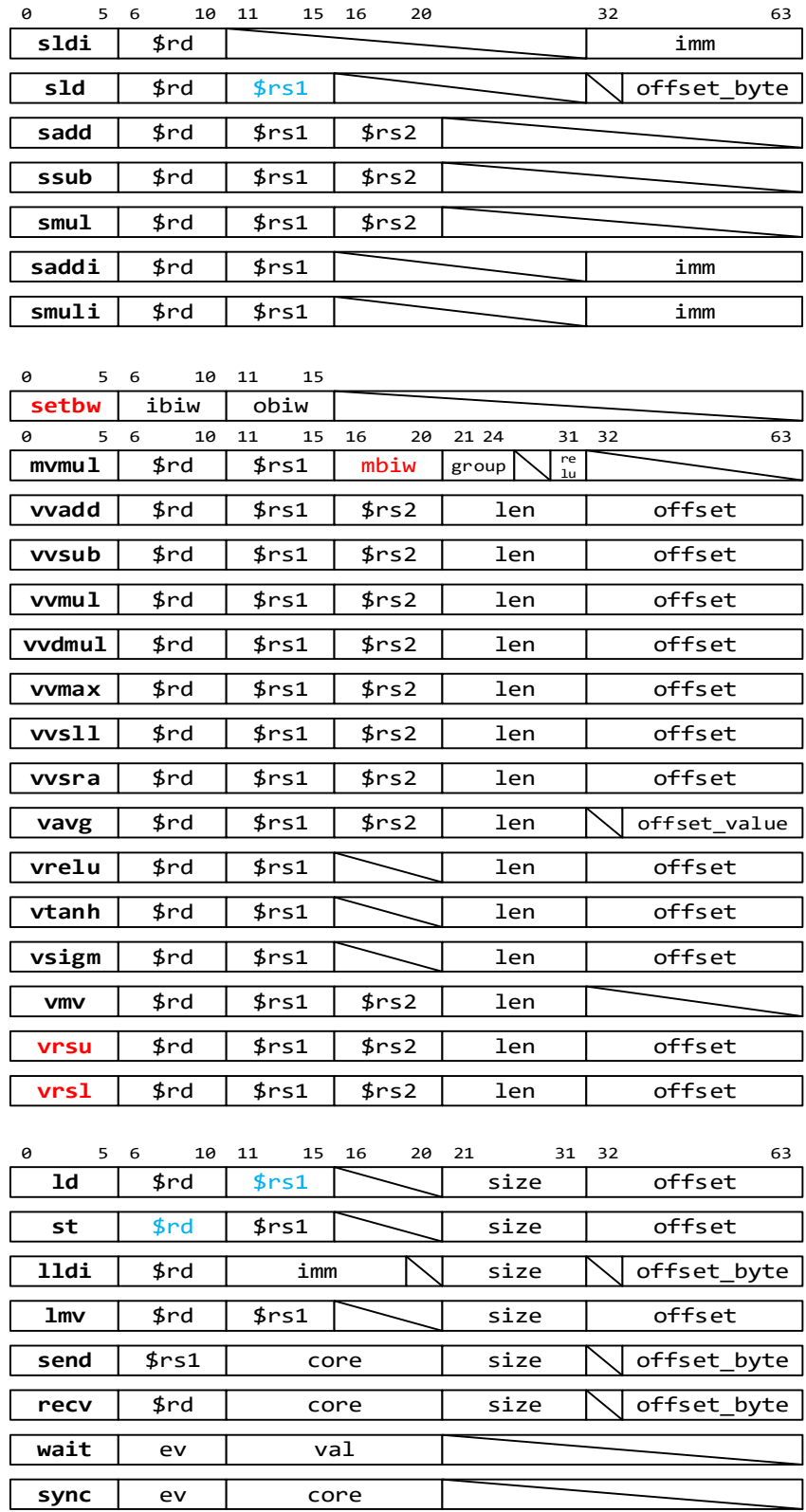

Figure 3: Instructions summary.